# Scale-free dynamics of somatic adaptability in immune system

Shiro Saito · Osamu Narikiyo

**Abstract**

The long-time dynamics of somatic adaptability in immune system is simulated by a simple physical model. The immune system described by the model exhibits a scale free behavior as is observed in living systems. The balance between the positive and negative feedbacks of the model leads to a robust immune system where the positive one corresponds to the formation of memory cells and the negative one to immunosuppression. Also the immunosenescence of the system is discussed based on the time-dependence of the epigenetic landscape of the adaptive immune cells in the shape space.

S. Saito · O. Narikiyo

Department of Physics, Kyushu University,

6-10-1 Hakozaki, Fukuoka 812-8581, Japan

e-mail: narikiyo@phys.kyushu-u.ac.jp

# Introduction

Theoretical studies of the immune system have a long and diverse history. Among them the network model introduced by Jerne (1974) is one of the most popular models for physicists and has developed into many variants (Perelson and Weisbuch 1997). In these models the adaptive immune system is described as a dynamical network which evolves according as experienced attacks by antigens. In this paper we succeed to the development and study the long-time dynamics of the network based on a physical model which we propose.

Recently many biological phenomena have been analyzed in terms of network (Junker and Schreiber 2008) at various hierarchies. Especially the scale-free network (Barabasi and Albert 1999; Jeong et al. 2000; Albert et al. 2000; Albert and Barabasi 2002) has been attracting many attentions. One of the most important properties of the scale-free network is its robustness. The robustness is necessary for systems to survive and biologically significant factor.

In the studies of the adaptive immune system such a scale-free network is observed experimentally (Naumov et al. 2003) in the repertoire of memory T-cells. Theoretical explanation of the experiment has been already attempted (Ruskin and Burns 2006) by the small-world construction (Albert and Barabasi 2002) which is one of the scenarios leading to the scale-free network. In this paper we propose another scenario of the scale-free behavior in the adaptive immune system starting from a minimal model of the immune response against repeated attacks by antigens. Our minimal model describes the population dynamics of immune cells in the so-called shape space. Using this minimal model we perform long-time simulations and obtain the sequences of epigenetic landscapes in the shape space.

Although the minimal model leads to a scale-free behavior, it becomes fragile after long-time experience against repeated attacks by antigens. By a positive feedback of the model the cells are concentrates at the positions, which have been attacked by antigens, in the shape space so that the cells are absent at the other positions, which are security holes of immune response, since the total number of immune cells is limited to be finite. Such fragility corresponds to immunosenescence by aging in living systems. The immunosenescence has been discussed theoretically from the view point of highly optimized tolerance (Stromberg and Carlson 2006) which is one of the scenarios leading to the scale-free behavior.

Since the immunosenescence results from a positive feedback in the minimal model, we add a negative feedback to balance with it. This negative feedback corresponding to immunosuppression is implemented by the avalanche in the population of immune cells. In the study of non-equilibrium statistical physics the avalanche in the sand-pile model

(Bak et al. 1987, 1988) is known as the dynamics of self-organized criticality and is expected to be a universal description of a wide class of dynamical phenomena. The self-organized criticality is also one of the scenarios leading to the scale-free behavior. We identify the dynamical network of the adaptive immune system as a self-organized critical state. Our model reinforced with the avalanche feedback reduces the immunosenescence substantially. At the same time it becomes robust against the perturbation to the adaptive immune network. Such robustness is expected and desirable.

**Population Model**

As a minimal model for the adaptive immune system we propose a model of population dynamics of immune cells. The cells are characterized by the position in the so-called shape space. The position represents not only the shape of the antibodies but also physico-chemical properties. Hereafter we only employ 2-dimensional shape space for simplicity of the implementation. We do not distinguish the species (naïve, memory, B-, T-, etc.) of the cells but characterize the cells by their lifetime. We introduce an abstract cell and the lifetime of the cell is prolonged if it experiences the attack by antigens. We assume that the lifetime is determined by the state of the network among immune cells but do not explicitly construct the network and treat it as the background that allows the change of the lifetime. Thus each cell in the adaptive immune network is characterized by the position in the shape space and by the lifetime in our minimal model.

Based on above assumptions we simulate the population dynamics of immune cells by the following rule. ( ⅰ ) We prepare a square lattice with $L \times L$ sites on which immune cells are distributed. Each site is represented by the coordinate $(i, j)$ with $i, j = 1, 2, 3, \cdots, L$. The population of the cell at the site $(i, j)$ and at the time $t$ is denoted as $z(i, j; t)$. The total number of immune cells is set as $N \times L \times L$ in the stationary state. ( ⅱ ) In the initial state $(N - N') \times L \times L$ cells are uniformly distributed and $N' \times L \times L$ cells are randomly distributed. ( ⅲ ) The time evolution of the adaptive immune system starts with the attack by an antigen to the site $(i, j)$ which is randomly chosen. The antigen introduced at the time $t$ is deleted if the condition

$$z(i, j; t) + z(i+1, j; t) + z(i-1, j; t) + z(i, j+1; t) + z(i, j-1; t) \geq Z \qquad (1)$$

is satisfied where the constant $Z$ is the threshold for the ability to delete the antigen. Otherwise the antigen remains at the site $(i, j)$ until the condition is satisfied. Here the summation over the team, the cell at $(i, j)$ and its nearest neighbor cells, takes into account the cooperative immune response of the underlying network. Then even if the cell at $(i, j)$ alone does not have enough ability, the team can respond to the antigen. The immune response to delete the antigen is completed at the next time step $t+1$. At the same time the population of the team is increased as

$$z(i, j; t+1) = z(i, j; t) + 2P$$
$$z(i \pm 1, j; t+1) = z(i \pm 1, j; t) + P \quad (2)$$
$$z(i, j \pm 1; t+1) = z(i, j \pm 1; t) + P$$

with the constant $P$. This increase reflects the formation of memory cells and the adaptive strengthen of the related part of the network. (ⅳ) During the next $6P$ time steps randomly chosen $6P$ cells are forced to die according to the rate for death $d(k,l;t)$. Consequently the total number of the immune cells becomes back to the value $N \times L \times L$ at the stationary state. The rate is defined by the inverse of the population $z(k,l;t)^{-1}$ as

$$d(k,l;t) = \frac{z(k,l;t)^{-1}}{\sum_{k'}\sum_{l'} z(k',l';t)^{-1}} \quad (3)$$

where $z(k,l;t)^{-1}$ is only assigned to the site with $z(k,l;t) \neq 0$. Thus the lifetime of the cell is set to be longer at the site with larger population. (ⅴ) During the next $Q$ time steps naïve cells are supplied to randomly chosen $Q$ sites. At each time step a naïve cell is added at randomly chosen site $(i, j)$ as

$$z(i, j; t+1) = z(i, j; t) + 1 \quad (4)$$

and a cell is removed from randomly chosen site $(k,l)$ according to the rate $d(k,l;t)$ as

$$z(k,l; t+1) = z(k,l; t) - 1 \quad (5)$$

to keep the total number of cells being constant. (ⅵ) After these $Q$ time steps interval we introduce the next attack by an antigen to a randomly chosen site and repeat the procedures, (ⅲ), (ⅳ), and (ⅴ), sequentially. The number of the repeated cycle is denoted by $A$ which is also the total number of antigens experienced by the adaptive immune system.

In the following the parameters are chosen as $(Z, P, Q) = (3, 10, 1000)$.

In Fig. 1 the distribution of the population is shown. In the region of relatively large population the distribution is scale-free; it shows a power-law behavior as is observed in living systems (Naumov et al. 2003). A scale-free behavior is expected from our rule of the rate for death which has some resemblance to the rule of the small-world construction (Albert and Barabasi 2002) which is one of the scenarios leading to the scale-free network. In the small-world case a new link is attached to a node with the probability which is proportional to the number of links already attached to it. Thus a positive feedback which favors the hubs with many links is implemented. Our rule with the control of lifetimes also favors the cells with large populations. Such a positive feedback magnifies the difference between the rich and the poor.

In the region of relatively small population it deviates from the power-law, because the total number of immune cells is limited to be finite and is short to link all the cells together as a scale-free network. Then cells with small population which do not respond against antigens are separated from the part of the network with large populations. Namely the deviation results from the lack of the resource. The finiteness of the resource is not taken into account in the case of small-world construction (Albert and Barabasi 2002).

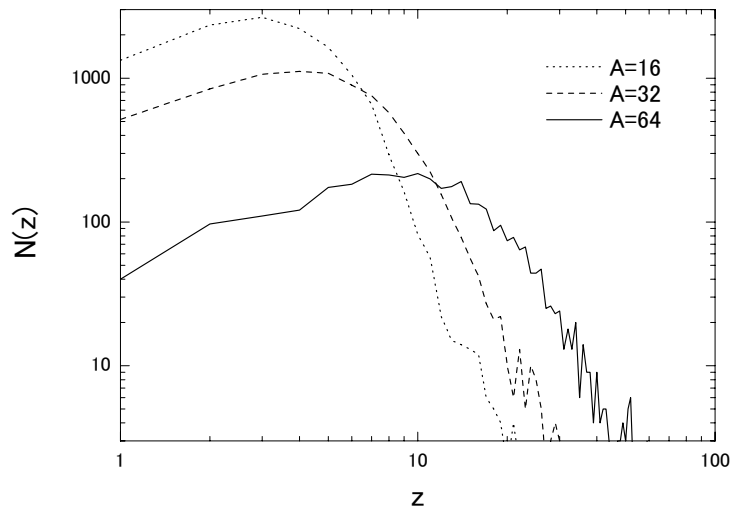

**Fig. 1**  The number of sites $N(z)$ with the population $z$ for $L = 128$ and $A = 16, 32, 64$.

A mean-field approximation to the small-world construction (Albert and Barabasi 2002) leads to the power of $-3$. The absolute values of the power seen in Fig. 1 are larger than 3 corresponding to the fact that the growth of the cells with the largest populations is slower than that of the number of links attached to hubs in the small-world construction. The growth in our case is relative and indirect which is the consequence of the survival, while the hubs grow by the direct increase of links. The growth by the indirect mechanism is slower than that by the direct one. The absolute values of the power become smaller as $A$ is increased. The reason is discussed in terms of the epigenetic landscape later.

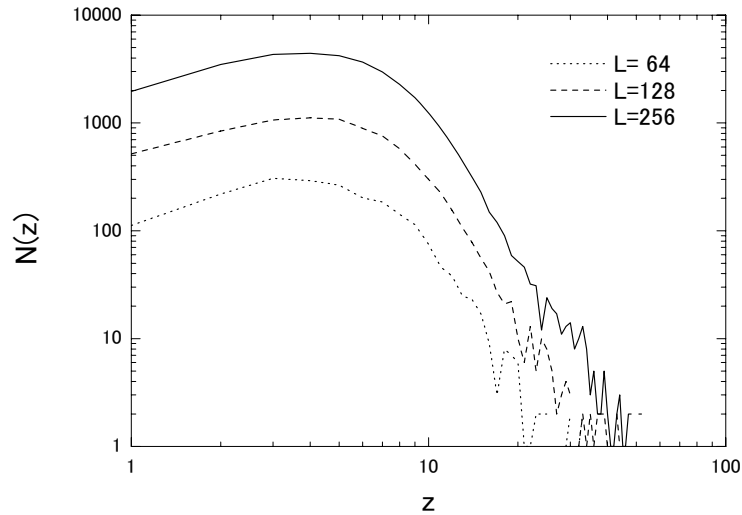

**Fig. 2**   The number of sites $N(z)$ with the population $z$ for $(L, A) = (128, 32), (256, 128), (512, 512)$.

  In Fig. 2 the distribution of the population is shown for various system sizes at the same $A$. The powers depicted by the straight lines are nearly the same and independent of the system size in the region of these simulations.

  In Fig. 3 the active site which is able to respond to the attack by antigens is plotted by the dot. As $A$ is increased, the relatively uniform distribution at earlier stage changes into a scattered distribution of clusters at later stage. Since the population in the cluster increases by the positive feedback, the total area covered by the clusters, namely the total number of active sites, decreases due to the finiteness of the resource as shown in Fig. 4. This decrease corresponds to the immunosenescence by aging in living systems. Figure 3 is the epigenetic landscape and Fig. 4 describes quantitatively the progress of the immunosenescence. The change in the power seen in Fig. 1 depending on $A$ is due to the merger of clusters which accelerate the growth of the cell populations in the survived cluster.

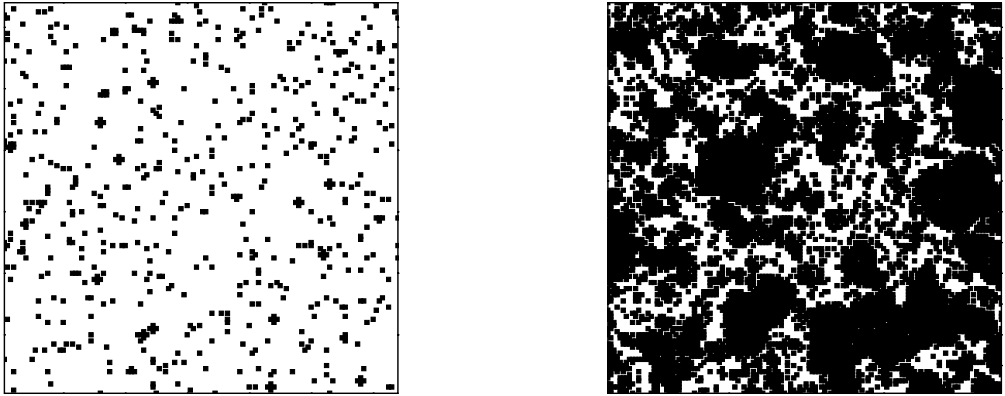

**Fig. 3**  The epigenetic landscape in the shape space where the site which satisfies Eq. (1) and is able to respond the antigen is depicted by the square dot for $(L, A) = (128, 64)$. The left sparse landscape is obtained for the population model. The right dense landscape is for the avalanche model.

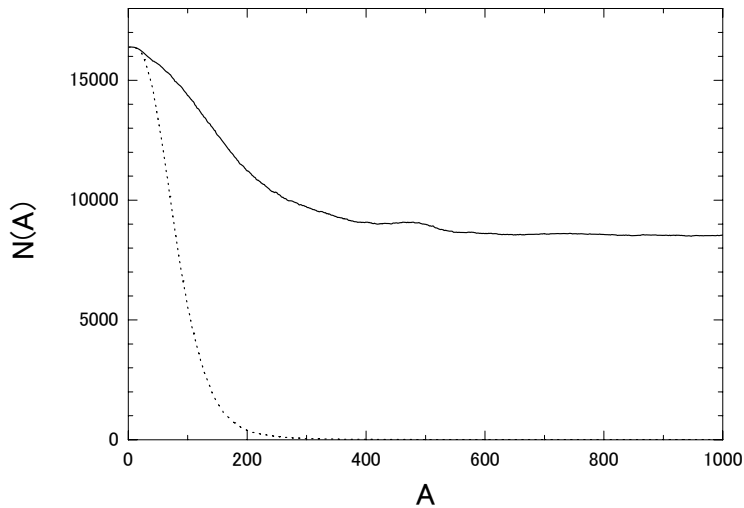

**Fig. 4**  The number of active sites $N(A)$, which satisfy Eq. (1) and can respond the antigen, after $A$ times of attacks for $L = 128$. The solid line is obtained for the avalanche model. The broken line is for the population model.

The immunosenescence has been discussed theoretically from the view point of highly optimized tolerance (Stromberg and Carlson 2006) which is one of the scenarios leading to the scale-free behavior. It is the fragility caused by the positive feedback and the lack of resources. In our model the fragility is severe and the active sites decreases substantially after the experience of repeated attacks by antigens. Then in order to construct a robust immune system we introduce a negative feedback, which balances with the positive one, in the following improved model.

**Avalanche Model**

Our population model only takes a positive feedback into account to obtain a minimal model. Such a feedback magnifies the difference between the rich and the poor and leads to a fragile state with heavy immunosenescence where the oligopoly of the finite resource by the rich results in unbalanced distribution of the cell population. In order to obtain a robust immune system we introduce a negative feedback to redistribute the resource to the poor. This negative feedback implements the immunosuppression in living systems.

We implement the redistribution by making avalanches among the populations. The dynamics of the interaction among cells are represented by avalanches. It is known that the avalanches in the sand-pile model (Bak et al. 1987, 1988) lead to the state of self-organized criticality with a scale-free distribution of the sizes of avalanches. Such a model is expected to be a universal description of a wide class of dynamical phenomena. We identify the dynamical network of the adaptive immune system as a self-organized critical state.

In the following we focus on the dynamics of the interaction network, while the population model mainly deals with stationary property.

The above simulation rule is modified by adding the following procedure at each time step of ($\vee$). If there is an excess site as $z(i, j;t) \geq C$ where $C$ is the threshold constant ($C \geq 4$), the avalanche around it is introduced so that the populations are redistributed as

$$z(i, j;t) \to z(i, j;t) - 4$$
$$z(i \pm 1, j;t) \to z(i \pm 1, j;t) + 1 \qquad (6)$$
$$z(i, j \pm 1;t) \to z(i, j \pm 1;t) + 1.$$

This procedure is continued within the time step until the excess site disappears. In general several avalanches, isolated or series of, occur within the time step.

In the following the parameters are chosen as $(Z, P, Q, C) = (3, 10, 1000, 8)$.

As seen in Fig. 4 the immunosenescence is substantially suppressed by the negative feedback.

In Fig. 5 the distribution of the area of the avalanche $S$ is shown. The distribution is scale-free; it shows a power-law behavior. A scale-free behavior is expected from our rule for the avalanche which is the same as that in the sand-pile model (Bak et al. 1987, 1988) of self-organized criticality which is one of the scenarios leading to the scale-free network. In the region of the largest avalanches it deviates from the power-law, because the size of the avalanche is cut off by the finite size of the shape space. At the same time the larger avalanche is harder to occur because of the lack of the resource.

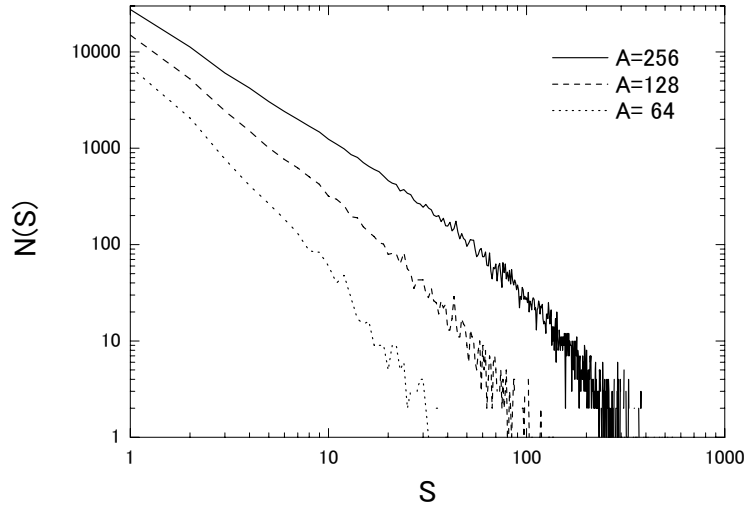

**Fig. 5** The number of avalanche events $N(S)$ with the avalanche size $S$ for $L = 128$ and $A = 64, 128, 256$. $N(S)$ counts all the avalanche events during $A$ times of attacks. $S$ is the number of sites involved in an avalanche event.

In Fig. 6 the distribution of the area is shown for various system sizes at the same $A$. The powers depicted by the straight lines are nearly the same and independent of the system size in the region of these simulations.

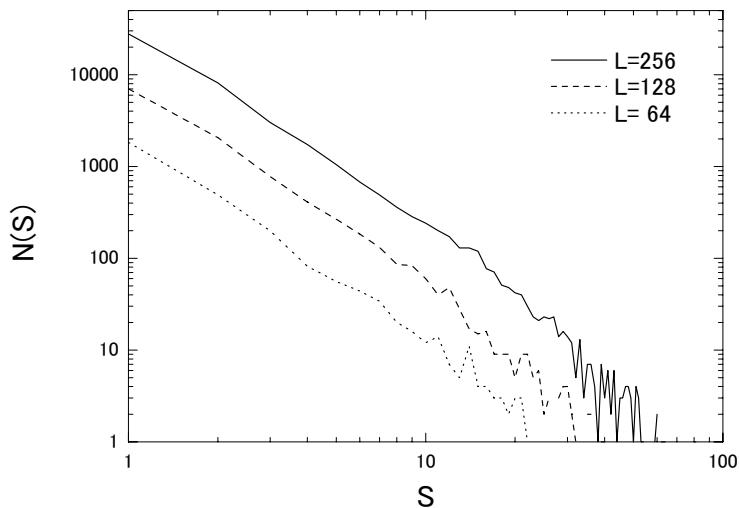

**Fig. 6**  The number of avalanche events $N(S)$ with the avalanche size $S$ for $(L, A) = (64, 16), (128, 64), (256, 256)$. $N(S)$ counts all the avalanche events during $A$ times of attacks. $S$ is the number of sites involved in an avalanche event.

In the sand-pile model (Bak et al. 1987, 1988) the absolute value of the power is about 1. The value in Figs. 5 and 6 is larger than 1. This difference comes from the finiteness of the resource not taken into account in the sand-pile model. The larger avalanche is harder to occur in our model because of the lack of the resource.

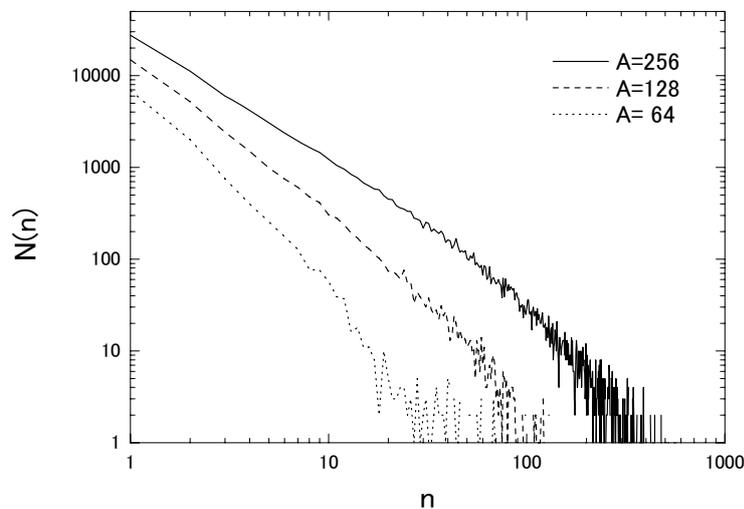

**Fig. 7**  The number of sites $N(n)$ which experienced $n$ times avalanche events for $L = 128$ and $A = 64, 128, 256$. $N(n)$ is the total number of experiences during $A$ times of attacks.

In Fig. 7 the distribution in terms of the number of events of avalanches in a time step is shown. It is the number of interactions among cells and represents the activity of the underlying dynamical network. This activity also exhibits a scale-free behavior.

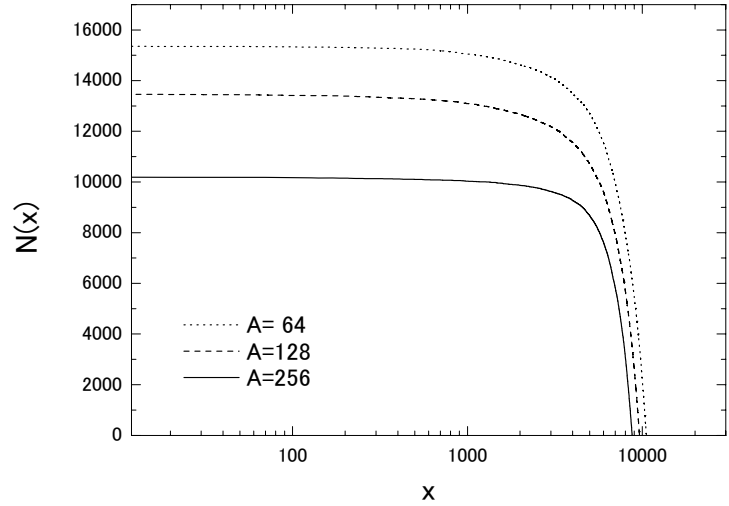

**Fig. 8** The number of active sites $N(x)$, which satisfy Eq. (1) and can respond the antigen, after $x$ destruction events for $L = 128$ and $A = 64, 128, 256$. The destruction events are introduced into the system which experienced $A$ times of attacks. A single destruction event is defined as the procedure $z(i, j; t) \to 0$ at a randomly chosen site where $z(i, j; t) \neq 0$.

In Fig. 8 we show the robustness of our adaptive immune network against the perturbation destroying the elements. It has been shown that the scale-free network by the small-world construction (Albert and Barabasi 2002) is highly tolerant to such a perturbation and robust. Our scale-free network is also tolerant and robust.

## Conclusion

We have introduced a simple physical model to describe the long-time dynamics of somatic adaptability in immune system. The population model, the tentative version of the model, exhibits a scale free behavior as is observed in living systems. However, the population model leads to a fragile state with heavy immunosenescence. This immunosenescence is the consequence of the positive feedback of the model. The positive feedback corresponds to the formation of memory cells in living systems and has a similar effect as in the case of the small-world network. In order to construct a robust system we have introduced a negative feedback which balances with the positive one. The negative feedback corresponds to the immunosuppression in living systems and is implemented by the avalanche among the populations. The dynamics of the interaction network represented by the avalanche exhibits a scale free behavior similar to the self-organized criticality of the sand-pile model. The avalanche model leads to a robust state which heavily reduces the immunosenescence and becomes tolerant against perturbations. The time-dependence of the immunosenescence is visualized as the epigenetic landscape in the shape space.

Although our model is highly abstract, we expect that it supplies a simple tool for long-time simulation and captures important features of the dynamics of adaptive immune system.